\makeatletter

\font\tenmsx=msxm10
\font\sevenmsx=msxm7
\font\fivemsx=msxm5
\font\tenmsy=msym10
\font\sevenmsy=msym7
\font\fivemsy=msym5
\newfam\msxfam
\newfam\msyfam
\textfont\msxfam=\tenmsx  \scriptfont\msxfam=\sevenmsx
  \scriptscriptfont\msxfam=\fivemsx
\textfont\msyfam=\tenmsy  \scriptfont\msyfam=\sevenmsy
  \scriptscriptfont\msyfam=\fivemsy

\def\hexnumber@#1{\ifnum#1<10 \number#1\else
 \ifnum#1=10 A\else\ifnum#1=11 B\else\ifnum#1=12 C\else
 \ifnum#1=13 D\else\ifnum#1=14 E\else\ifnum#1=15 F\fi\fi\fi\fi\fi\fi\fi}

\def\msx@{\hexnumber@\msxfam}
\def\msy@{\hexnumber@\msyfam}
\mathchardef\boxdot="2\msx@00
\mathchardef\boxplus="2\msx@01
\mathchardef\boxtimes="2\msx@02
\mathchardef\square="0\msx@03
\mathchardef\blacksquare="0\msx@04
\mathchardef\centerdot="2\msx@05
\mathchardef\lozenge="0\msx@06
\mathchardef\blacklozenge="0\msx@07
\mathchardef\circlearrowright="3\msx@08
\mathchardef\circlearrowleft="3\msx@09
\mathchardef\rightleftharpoons="3\msx@0A
\mathchardef\leftrightharpoons="3\msx@0B
\mathchardef\boxminus="2\msx@0C
\mathchardef\Vdash="3\msx@0D
\mathchardef\Vvdash="3\msx@0E
\mathchardef\vDash="3\msx@0F
\mathchardef\twoheadrightarrow="3\msx@10
\mathchardef\twoheadleftarrow="3\msx@11
\mathchardef\leftleftarrows="3\msx@12
\mathchardef\rightrightarrows="3\msx@13
\mathchardef\upuparrows="3\msx@14
\mathchardef\downdownarrows="3\msx@15
\mathchardef\upharpoonright="3\msx@16

\mathchardef\downharpoonright="3\msx@17
\mathchardef\upharpoonleft="3\msx@18
\mathchardef\downharpoonleft="3\msx@19
\mathchardef\rightarrowtail="3\msx@1A
\mathchardef\leftarrowtail="3\msx@1B
\mathchardef\leftrightarrows="3\msx@1C
\mathchardef\rightleftarrows="3\msx@1D
\mathchardef\Lsh="3\msx@1E
\mathchardef\Rsh="3\msx@1F
\mathchardef\rightsquigarrow="3\msx@20
\mathchardef\leftrightsquigarrow="3\msx@21
\mathchardef\looparrowleft="3\msx@22
\mathchardef\looparrowright="3\msx@23
\mathchardef\circeq="3\msx@24
\mathchardef\succsim="3\msx@25
\mathchardef\gtrsim="3\msx@26
\mathchardef\gtrapprox="3\msx@27
\mathchardef\multimap="3\msx@28
\mathchardef\therefore="3\msx@29
\mathchardef\because="3\msx@2A
\mathchardef\doteqdot="3\msx@2B

\mathchardef\triangleq="3\msx@2C
\mathchardef\precsim="3\msx@2D
\mathchardef\lesssim="3\msx@2E
\mathchardef\lessapprox="3\msx@2F
\mathchardef\eqslantless="3\msx@30
\mathchardef\eqslantgtr="3\msx@31
\mathchardef\curlyeqprec="3\msx@32
\mathchardef\curlyeqsucc="3\msx@33
\mathchardef\preccurlyeq="3\msx@34
\mathchardef\leqq="3\msx@35
\mathchardef\leqslant="3\msx@36
\mathchardef\lessgtr="3\msx@37
\mathchardef\backprime="0\msx@38
\mathchardef\risingdotseq="3\msx@3A
\mathchardef\fallingdotseq="3\msx@3B
\mathchardef\succcurlyeq="3\msx@3C
\mathchardef\geqq="3\msx@3D
\mathchardef\geqslant="3\msx@3E
\mathchardef\gtrless="3\msx@3F
\mathchardef\sqsubset="3\msx@40
\mathchardef\sqsupset="3\msx@41
\mathchardef\trianglerighteq="3\msx@44
\mathchardef\trianglelefteq="3\msx@45
\mathchardef\bigstar="0\msx@46
\mathchardef\between="3\msx@47
\mathchardef\blacktriangledown="0\msx@48
\mathchardef\blacktriangleright="3\msx@49
\mathchardef\blacktriangleleft="3\msx@4A
\mathchardef\blacktriangle="0\msx@4E
\mathchardef\triangledown="0\msx@4F
\mathchardef\eqcirc="3\msx@50
\mathchardef\lesseqgtr="3\msx@51
\mathchardef\gtreqless="3\msx@52
\mathchardef\lesseqqgtr="3\msx@53
\mathchardef\gtreqqless="3\msx@54
\mathchardef\Rrightarrow="3\msx@56
\mathchardef\Lleftarrow="3\msx@57
\mathchardef\veebar="2\msx@59
\mathchardef\barwedge="2\msx@5A
\mathchardef\doublebarwedge="2\msx@5B
\mathchardef\angle="0\msx@5C
\mathchardef\measuredangle="0\msx@5D
\mathchardef\sphericalangle="0\msx@5E
\mathchardef\varpropto="3\msx@5F
\mathchardef\smallsmile="3\msx@60
\mathchardef\smallfrown="3\msx@61
\mathchardef\Subset="3\msx@62
\mathchardef\Supset="3\msx@63
\mathchardef\Cup="2\msx@64

\mathchardef\Cap="2\msx@65

\mathchardef\curlywedge="2\msx@66
\mathchardef\curlyvee="2\msx@67
\mathchardef\leftthreetimes="2\msx@68
\mathchardef\rightthreetimes="2\msx@69
\mathchardef\subseteqq="3\msx@6A
\mathchardef\supseteqq="3\msx@6B
\mathchardef\bumpeq="3\msx@6C
\mathchardef\Bumpeq="3\msx@6D
\mathchardef\lll="3\msx@6E

\mathchardef\ggg="3\msx@6F

\mathchardef\circledS="0\msx@73
\mathchardef\pitchfork="3\msx@74
\mathchardef\dotplus="2\msx@75
\mathchardef\backsim="3\msx@76
\mathchardef\backsimeq="3\msx@77
\mathchardef\complement="0\msx@7B
\mathchardef\intercal="2\msx@7C
\mathchardef\circledcirc="2\msx@7D
\mathchardef\circledast="2\msx@7E
\mathchardef\circleddash="2\msx@7F
\def\ulcorner{\delimiter"4\msx@70\msx@70 }
\def\urcorner{\delimiter"5\msx@71\msx@71 }
\def\llcorner{\delimiter"4\msx@78\msx@78 }
\def\lrcorner{\delimiter"5\msx@79\msx@79 }
\def\yen{\mathhexbox\msx@55 }
\def\checkmark{\mathhexbox\msx@58 }
\def\circledR{\mathhexbox\msx@72 }
\def\maltese{\mathhexbox\msx@7A }
\mathchardef\lvertneqq="3\msy@00
\mathchardef\gvertneqq="3\msy@01
\mathchardef\nleq="3\msy@02
\mathchardef\ngeq="3\msy@03
\mathchardef\nless="3\msy@04
\mathchardef\ngtr="3\msy@05
\mathchardef\nprec="3\msy@06
\mathchardef\nsucc="3\msy@07
\mathchardef\lneqq="3\msy@08
\mathchardef\gneqq="3\msy@09
\mathchardef\nleqslant="3\msy@0A
\mathchardef\ngeqslant="3\msy@0B
\mathchardef\lneq="3\msy@0C
\mathchardef\gneq="3\msy@0D
\mathchardef\npreceq="3\msy@0E
\mathchardef\nsucceq="3\msy@0F
\mathchardef\precnsim="3\msy@10
\mathchardef\succnsim="3\msy@11
\mathchardef\lnsim="3\msy@12
\mathchardef\gnsim="3\msy@13
\mathchardef\nleqq="3\msy@14
\mathchardef\ngeqq="3\msy@15
\mathchardef\precneqq="3\msy@16
\mathchardef\succneqq="3\msy@17
\mathchardef\precnapprox="3\msy@18
\mathchardef\succnapprox="3\msy@19
\mathchardef\lnapprox="3\msy@1A
\mathchardef\gnapprox="3\msy@1B
\mathchardef\nsim="3\msy@1C
\mathchardef\napprox="3\msy@1D
\mathchardef\nsubseteqq="3\msy@22
\mathchardef\nsupseteqq="3\msy@23
\mathchardef\subsetneqq="3\msy@24
\mathchardef\supsetneqq="3\msy@25
\mathchardef\subsetneq="3\msy@28
\mathchardef\supsetneq="3\msy@29
\mathchardef\nsubseteq="3\msy@2A
\mathchardef\nsupseteq="3\msy@2B
\mathchardef\nparallel="3\msy@2C
\mathchardef\nmid="3\msy@2D
\mathchardef\nshortmid="3\msy@2E
\mathchardef\nshortparallel="3\msy@2F
\mathchardef\nvdash="3\msy@30
\mathchardef\nVdash="3\msy@31
\mathchardef\nvDash="3\msy@32
\mathchardef\nVDash="3\msy@33
\mathchardef\ntrianglerighteq="3\msy@34
\mathchardef\ntrianglelefteq="3\msy@35
\mathchardef\ntriangleleft="3\msy@36
\mathchardef\ntriangleright="3\msy@37
\mathchardef\nleftarrow="3\msy@38
\mathchardef\nrightarrow="3\msy@39
\mathchardef\nLeftarrow="3\msy@3A
\mathchardef\nRightarrow="3\msy@3B
\mathchardef\nLeftrightarrow="3\msy@3C
\mathchardef\nleftrightarrow="3\msy@3D
\mathchardef\divideontimes="2\msy@3E
\mathchardef\varnothing="0\msy@3F
\mathchardef\nexists="0\msy@40
\mathchardef\mho="0\msy@66
\mathchardef\thorn="0\msy@67
\mathchardef\beth="0\msy@69
\mathchardef\gimel="0\msy@6A
\mathchardef\daleth="0\msy@6B
\mathchardef\lessdot="3\msy@6C
\mathchardef\gtrdot="3\msy@6D
\mathchardef\ltimes="2\msy@6E
\mathchardef\rtimes="2\msy@6F
\mathchardef\shortmid="3\msy@70
\mathchardef\shortparallel="3\msy@71
\mathchardef\smallsetminus="2\msy@72
\mathchardef\thicksim="3\msy@73
\mathchardef\thickapprox="3\msy@74
\mathchardef\approxeq="3\msy@75
\mathchardef\succapprox="3\msy@76
\mathchardef\precapprox="3\msy@77
\mathchardef\curvearrowleft="3\msy@78
\mathchardef\curvearrowright="3\msy@79
\mathchardef\digamma="0\msy@7A
\mathchardef\varkappa="0\msy@7B
\mathchardef\hslash="0\msy@7D
\mathchardef\hbar="0\msy@7E
\mathchardef\backepsilon="3\msy@7F
\def\Bbb{\ifmmode\let\next\Bbb@\else
 \def\next{\errmessage{Use \string\Bbb\space only in math mode}}\fi\next}
\def\Bbb@#1{{\Bbb@@{#1}}}
\def\Bbb@@#1{\fam\msyfam#1}

\def\inv{^{\raise.15ex\hbox{${
  \scriptscriptstyle -}$}\kern-.05em 1}}

\def\Dsl{\,\raise.15ex\hbox{$/$}\mkern-13.5mu D}
\def\dsl{\raise.15ex\hbox{$/$}\kern-.57em\hbox{$\partial$}}

\def\lspace{\ifx\answ\bigans{}\else\qquad\fi}

\def\CR{\hbox{{$\cal R$}}}

\def\lform{\hbox{$\sqcup$}\llap{\hbox{$\sqcap$}}}
\def\darr#1{\raise1.5ex\hbox{$\leftrightarrow$}
\mkern-16.5mu #1}
\def\INT{{\textstyle \int\kern-.642em\int}}

\def\C{{\Bbb C}}

\def\eps{{\epsilon}}

\def\small{\scriptstyle}
\def\tens{\mathop{\otimes}}

\def\la{{\triangleright}}\def\ra{{\triangleleft}}

\def\isom{{\cong}}

\def\eqn#1#2{\begin{equation}#2\label{#1}\end{equation}}
\def\o{{}_{(1)}}\def\t{{}_{(2)}}

\def\und#1{{\underline {#1}}}

\def\uo{{{}^{(1)}}}\def\ut{{{}^{(2)}}}

\def\text#1{\mbox{\rm #1}}
\def\note#1{}

\def\blacksquare{{\lform}}%AMS Tex Fakes
\def\frac#1#2{{{#1\over#2}}}

\def\und#1{{\underline{#1}}}

\def\<{\langle}
\def\>{\rangle}

\makeatother

\def\tr{{\rm tr}}

\documentstyle[12pt]{article}
\begin{document}
\newtheorem{prop}{Proposition}[section]
\newtheorem{lemma}[prop]{Lemma}
\newtheorem{thm}[prop]{Theorem}
\newtheorem{df}[prop]{Definition}
\newtheorem{rk}[prop]{Remark}
\newtheorem{cor}[prop]{Corollary}
\newtheorem{ex}[prop]{Example}
\baselineskip 13pt
\title{\marginpar{\vskip -.5in \hskip -1in \small DAMTP/92-51}
{\rm Quantum Group Gauge Theory\\ on Classical Spaces}}

\author{Tomasz Brzezi\'{n}ski \thanks{Supported by St. John's College,
Cambridge \& KBN grant 2 0218 91 01} \& Shahn Majid \thanks{SERC
Fellow and Drapers Fellow of Pembroke College, Cambridge}\\ \\
Department of Applied Mathematics \& Theoretical Physics \\
University of Cambridge, CB3 9EW, U.K.}
\date{August 1992}

\maketitle

\begin{quote}ABSTRACT We study the quantum group gauge theory developed
elsewhere in the limit when the base space (spacetime) is a classical
space rather than a general quantum space. We show that this limit of
the theory for gauge quantum group $U_q(g)$ is isomorphic to usual
gauge theory with Lie algebra $g$. Thus a new kind of gauge theory is
not obtained in this way, although we do find some differences in the
coupling to matter. Our analysis also illuminates certain
inconsistencies in previous work on this topic where a different
conclusion had been reached. In particular, we show that the use of the
quantum trace in defining a Yang-Mills action in this setting as
claimed in \cite{wu1}\cite{isaev1} is not appropriate.
\end{quote} \baselineskip 20pt

1. Recently a number of physicists have tried to develop a $q$-analogue
or quantum-group gauge theory in which the gauge fields have values in
$U_q(g)$ but spacetime is an ordinary manifold
\cite{arefeva1}\cite{wu1}\cite{isaev1}, \cite{hirayama1}. Here we want
to clarify some of the difficulties that arise in this context in the
light of our own general quantum group gauge theory developed in
\cite{brzezinski6}. In our work both the gauge group and spacetime were
allowed to be quantum spaces and non-trivial examples constructed. We now
study the limit of this theory in the case when the base space becomes a usual
classical one and show that  the resulting gauge theory in this case
is necessarily isomorphic to usual gauge theory at least at the level of
the gauge fields. There can nevertheless
be some subtle differences when the coupling to matter fields is
considered. We then contrast our result with some of the
above-mentioned previous literature, concentrating on
\cite{wu1}\cite{isaev1}. We show that the Yang-Mills action proposed in
these papers (via the quantum trace) is not fully gauge-invariant. We
also point out that a reasonable interpretation of some of the formulae
in \cite{wu1} is not possible unless the base space is given by the
quantum group itself, which is unfortunately not appropriate when the
base is classical as assumed. The paper concludes with a discussion of the
status of the matter fields in our theory.

In our theory \cite{brzezinski6} the role of gauge group is played by a
quantum group function algebra $A$  and base space by the (quantum)
algebra of functions $B$ on spacetime. $A$ has a standard Hopf algebra
structure \cite{abe1} \cite{sweedler1} given by the comultiplication
$\Delta$, counit $\epsilon$, and the antipode $S$. For trivial bundles,
a gauge field is a map $\beta: A\to \Omega ^1(B)$, such that $\beta (1)
=0$, where $\Omega^1(B)$ is the space of differential one-forms on
$B$.  A gauge transformation is a {\em convolution-invertible} map
$\gamma :  A \rightarrow B$ such that $\gamma (1_A) =  1_B$. Here, if
$\gamma,\gamma':A\to B$ are two linear maps then their convolution
product is given as another map $\gamma*\gamma':A\to B$ by
\begin{equation}
\gamma*\gamma'=\cdot_B(\gamma\tens \gamma')\Delta\end{equation}
where $\cdot_B$ is the product in $B$. This defines the
convolution algebra of maps, with identity provided by the counit
$\eps:A\to \C$ viewed as a map $A\to B$ by composing with $1_B$. We
require our gauge transformations to be invertible elements in this
convolution algebra. This technical invertibility requirement will have
important consequences in what follows.

Next, the fields on which our covariant derivative acts are linear maps
$\sigma: V\to B$, $\sigma (1_V) = 1_B$ where $V$ is a quantum algebra
of functions on a vector space (for example, the quantum plane) on
which $A$ coacts i.e.  $V$ is a space equipped with a corepresentation
$\rho:V\to A\tens V$. This is such that $(id \otimes \rho)\rho =
(\Delta \otimes id) \rho$ and $(\epsilon \otimes id)\rho = id$.
Moreover we assume that $\rho$ is an algebra homomorphism. In this case
we can extend our convolution product above to the product of maps
$\sigma: V  \rightarrow B$ and $\gamma: A \rightarrow B$ given as a map
$\gamma *  \sigma  : V \rightarrow B$ by
\begin{equation} \gamma *
\sigma= \cdot_B(\gamma\tens\sigma)\rho.  \label{convtrans}\end{equation}
Similarly we
can define $\beta*\sigma:A\to\Omega^1(B)$ using now the product of
$\Omega^1(B)$ with $B$ rather than $\cdot_B$. Finally, we can define
$\beta*\beta$ using in place of $\cdot_B$ the product of $\Omega^1(B)$
with itself to give an element of $\Omega^2(B)$ (this can be done
either with a wedge product in the exterior algebra or, more generally,
in the differential algebra $\Omega(B)$). With this notation, the
formulae for the covariant derivative $\nabla$ corresponding to the
gauge potential $\beta$ and the curvature $F:  A \rightarrow \Omega
^2(B)$ read \cite{brzezinski6} \begin{equation} \nabla \sigma  =
d\sigma + \beta * \sigma \end{equation} \begin{equation} F = d\beta +
\beta * \beta, \end{equation} while the action of gauge transformations
on sections and gauge fields, and the resulting transformations of
$F,\nabla$ come out as \cite{brzezinski6} \begin{equation}
\sigma^\gamma=\gamma*\sigma \end{equation} \begin{equation}
 \beta^\gamma = \gamma *\beta  * \gamma^{-1}+ \gamma *d(\gamma^{-1})
 \label{beta.trans} \end{equation} \begin{equation} \nabla^\gamma  =
\gamma *\nabla\gamma^{-1} * \end{equation} \begin{equation} F^\gamma =
\gamma *  F * \gamma^{-1}.  \end{equation} Moreover one has the Bianchi
identity \begin{equation} dF + \beta  * F - F * \beta = 0.
\end{equation}

These local formulae  are obtained from
the general theory of quantum  principal and vector bundles
presented in \cite{brzezinski6}. The gauge potential $\beta$
corresponds to the local version of the connection 1-form and
the connection is defined as an invariant projection on the subspace
of the vertical one-forms on the total space of the bundle.

The formulae above simplify further and take more familiar form if we
consider $A$ a matrix quantum group \cite{frt1}. For example, if
$A=SL_q(2)$ and $V=\Bbb C_q$ (the quantum plane), the formulae appear
in matrix form as follows. Write $\gamma^i{}_j=\gamma(t^i{}_j)$,
$(\gamma^{-1})^i{}_j=\gamma^{-1}(t^i{}_j)$, $\beta^i{}_j =
\beta(t^i{}_j)$, $F^i{}_j = F(t^i{}_j)$, and $\sigma^i = \sigma(v^i)$
where $t^i{}_j$, $i,j =1,2$ are the generators of $SL_q(2)$ and $v^i$,
$i=1,2$ are the generators of ${\Bbb C_q}$.  Then for example, the
transformation of the gauge potential $\beta$, the curvature and its
transformation come out as \begin{equation} (\beta^\gamma)^i{}_j =
\gamma^i{}_k \beta^k{}_l (\gamma^{-1})^l{}_j + \gamma^i{}_k
d(\gamma^{-1})^k{}_j. \label{betamat.trans} \end{equation}
\begin{equation} F^i{}_j = d\beta^i{}_j + \beta^i{}_k\beta^k{}_j
\end{equation} \begin{equation} F^i{}_j =
\gamma^i{}_kF^k{}_l(\gamma^{-1})^l{}_j.  \end{equation}

The same formulae work quite generally with $A$ a matrix quantum group
\cite{frt1} and $V$ the Zamolodchikov algebra (for example). Note that
each entry $\gamma^i{}_j$ is an element of $B$, so if $B=C(X)$ for a
classical space-time $X$ then these entries are functions $x\to
\gamma^i{}_j(x)$ as usual. Each value $\gamma^i{}_j(x)$ is a $\Bbb
C$-number. On the other hand, a gauge transformation involves not only
$\gamma(t^i{}_j)$ but the value on all products such as
$\gamma^i{}_j{}^k{}_l=\gamma(t^i{}_j t^k{}_l)$ etc. Alternatively, one
can work abstractly with the generators $t^i{}_j$ rather than
emphasising any choice of $\gamma$.  We now specialise to this case
$B=C(X)$.  \vspace{12pt}

2. In the case when there is a usual space-time $X$ it is also
convenient to work with the quantum enveloping algebra $H$ (such
as $H=U_q(g)$) dual to $A$. We now dualise our formulae to this
case. Thus a gauge transformation is a suitably invertible map
$\gamma :X\to H$, a gauge field is an element $A\in \Omega^1(X)\tens H$
etc, where $\Omega^1(X)$ denotes the 1-forms or cotangent space to $X$. The
requirement of convolution invertibility of $\gamma$ now becomes
that at each point $x \in X$, $\gamma(x)$ is an invertible element of
$H$. Given this, the formulae are
\begin{equation}
 A^\gamma(x)= \gamma(x) A(x) (\gamma(x))^{-1}+ \gamma(x)
d(\gamma(x)^{-1})
\label{class1}
\end{equation}
\begin{equation}
F(x) = dA(x) + A(x) \wedge A(x)
\end{equation}
\begin{equation}
F^\gamma(x)=\gamma(x) F(x)(\gamma(x))^{-1}.
\label{class3}
\end{equation}

For example, one can write down at once a gauge-invariant Yang-Mills
action,

\begin{equation}L =-\int_X tr (F \wedge *F) \end{equation}
where $*$ denotes the Hodge $*$, as well as all other familiar
actions. Note that {\em we do not use the quantum trace} here
since only the ordinary trace is necessarily invariant under
conjugation by $\gamma$.

Thus, at least in the approach of \cite{brzezinski6} the quantum-group
gauge theory formulae in the case when the base is classical are just
analogous to the usual ones.

At first sight one can nevertheless envisage useful applications in
which $H$ is a non-standard quantum group, or perhaps a discrete group
algebra. There are many non-standard quantum groups (without familiar
classical limits) which surely offer here new possibilities for
phenomenology. Let us point out however, that this question needs to be
approached with care.  The reason is that the true gauge group above is
not really the quantum group $H$ but the ordinary group $H^{inv}$ of
invertible elements of $H$, for it is here that the gauge
transformations take their values.

This situation can be demonstrated in the case $H=U_q(g)$, i.e.
q-deformed universal enveloping algebra of Lie algebra $g$. As
an algebra it is known \cite{drinfeld5} that $U_q(g)\isom U(g)$
the usual enveloping algebra. Hence the invertible elements of
$U_q(g)$ are just the invertible elements of $U(g)$, i.e. the
usual group elements $\exp(\xi)$ for $\xi\in g$ (we work here
with topological Hopf algebras). Thus the gauge group in this
case is just the usual gauge group $G$ with Lie algebra $g$,
viewed in $U_q(g)$ via the isomorphism.  In this way, the
formulae (\ref{class1})-(\ref{class3}) are exactly isomorphic to
the usual formulae of gauge theory.

This is not to say that there are no differences at all with usual
gauge theory, particularly when one considers the matter fields
$\sigma$. We discuss this at the end of the paper.  \vspace{12pt}

3. Our conclusion so far is in marked contrast to claims in the
literature \cite{arefeva1}\cite{wu1}\cite{isaev1}\cite{hirayama1}. Here
we focus on \cite{wu1}\cite{isaev1} and explain some key
differences with our approach and some fundamental problems that arise
in these works as a result.

We begin by distinguishing carefully between two possibilities for how
to generalize the usual transformation of the curvature $F(x)\to
\gamma(x)F(x)(\gamma(x))^{-1}$. If one retains this pointwise
conjugation in the quantum case then we are in the situation above, and
we must use the classical trace in defining the Yang-Mills action as
explained. An alternative, which seems to be the one most relevant to
\cite{wu1}\cite[eqns (3.13)-(3.14)]{isaev1}, is to try to use the
quantum adjoint action, in which case one must use the quantum trace
(this is specifically  designed to be invariant under such a quantum
adjoint action). In our present setting where $\gamma(x),
F_{\mu\nu}\in U_q(g)$ this second possibility would mean that $F$
transforms as

\eqn{quadj}{ F\to F^\gamma=\sum \gamma\o F S\gamma\t}
where $\Delta \gamma= \sum \gamma\o \otimes \gamma\t$ is the
coproduct and $S$ the antipode of the quantum group. The
relevant quantum trace now is the category-theoretic trace as
used in defining the quantum dimension \cite[Section 7.5]{majid1},
namely
\begin{equation} \underline \tr(\ )=\tr((\ )u),
\qquad u=\sum (S\CR\ut)\CR\uo\label{trace}\end{equation}
where $\CR=\sum \CR\uo\otimes\CR\ut$ is the universal $R$-matrix
and $\tr$ is the usual trace.  The problem with this approach, as is
perhaps well-known, is that there is no way to set up the gauge
theory so that $F$ really transforms by the quantum adjoint
action. More precisely, one can allow the gauge potential $A$ to
transform by $A\to
\sum \gamma\o AS\gamma\t + \sum\gamma\o dS\gamma\t$ for
example, but as far as we know there is no way to define $F$
from this to transform as desired.  Note also that products of
$F$ etc as in the Yang-Mills action would have to transform as a
module-algebra, i.e.  $(FF)^\gamma=\sum F^{\gamma\o} F^{\gamma\t}$
for $\int_X dx\underline\tr F_{\mu\nu}(x)F^{\mu\nu}(x)$ etc to be
invariant.

In summary, we see some {\em a priori} problems in the attempt to use
the quantum trace as claimed in \cite{wu1}\cite{isaev1}.  If one wants
to use gauge transformations as in (\ref{class3}) then the use of the
quantum trace (discussed by \cite{wu1}\cite{isaev1} in a dual language)
rather than the usual trace, renders the action non-invariant. If on
the other hand one tries to use some kind of quantum adjoint action
like (\ref{quadj}) then it seems unlikely that there is a formulation
of transformation of gauge fields leading to the desired transformation
of the curvature.

Finally, we consider the situation explicitly in the matrix form that is
proposed in \cite[eqns (3.13)-(3.14)]{isaev1}. We show that the
$GL_q(2)$ invariant Yang-Mills lagrangian proposed there based on the
matrix quantum trace is not gauge invariant if we compose two (or more)
transformations. The same applies to a similar proposal in \cite{wu1}
for $SU_q(2)$. In both cases one uses the well-known quantum trace of a
$2\times 2$ matrix $E$ as
\begin{equation}
\tr_qE = q^{-1}E_{11}+qE_{22}.
\end{equation}
If $G_{ij}\in GL_q(2)$ is a matrix of generators then the
transformation
\begin{equation}E\rightarrow E'= GESG\end{equation}
leaves $\tr_qE$ invariant provided $[E_{ij},G_{kl}] = 0$, for all
$i,j,k,l =1,2$. This definition is used to the construction of
the invariant lagrangian
\begin{equation}{\cal L}_q = \tr_q(F_{\mu\nu}F^{\mu\nu})\end{equation}
where $F_{\mu\nu}$ are components of the gauge-field strength.
This lagrangian is indeed invariant under the transformation
\begin{equation}
F_{\mu\nu} \rightarrow F'_{\mu\nu} = GF_{\mu\nu}SG=F^I_{\mu\nu}\tens
G\sigma^I SG; \quad [F^I_{\mu\nu}, G_{ij}]=0
\label{curv.trans}
\end{equation}
where $\sigma^I$ are the identity matrix for $\sigma^0$ and the Pauli
matrices for $\sigma^1,\sigma^2,\sigma^3$. Of course, we can equally
well work with $F_{\mu\nu}{}_{ij}=F^I_{\mu\nu}(\sigma^I)_{ij}$, which
is the notation used in \cite[Section IV]{wu1}. The idea is that it
transforms like the $E_{ij}$ above. It does not matter here exactly
what kind of object the components $F^I_{\mu\nu}$ or
$F_{\mu\nu}{}_{ij}$ are as long as they all commute with all the
$G_{ij}$, in which case the defining property of the quantum trace
ensures that \begin{equation}\tr_q(F_{\mu\nu}F_{\mu\nu}) =
\tr_q(F'_{\mu\nu}F'_{\mu\nu}).\end{equation} Unfortunately, we cannot
in general repeat this gauge transformation because the matrix entries
of the transformed $F'_{\mu\nu}$ need not similarly commute with the
$G_{ij}$.  For example, even if $F_{\mu\nu}$ is a classical $\Bbb
C$-matrix (so that its matrix entries commute with the $G_{ij}$), the
transformed field strength $F'_{\mu\nu}$ involves $G, SG$ and hence
need not commute.

Another way to see this is that if we do go ahead and repeat the gauge
transformation, i.e. consider
\begin{equation}F'_{\mu\nu}\rightarrow F'{}'_{\mu\nu} =
G^2F_{\mu\nu}(SG^2)\end{equation} we see immediately that
\begin{equation}\tr_q(F'{}'_{\mu\nu}F'{}'_{\mu\nu}) \neq
\tr_q(F_{\mu\nu}F_{\mu\nu})\end{equation} if $q \neq 1$. This is because
\cite{corrigan1}\cite{wess1} the matrix $G'=G^2$ obeys the relations of
the quantum group $GL_{q^2}(2)$. Therefore the trace $\tr_q$ has to be
replaced by
$\tr_{q^2}$ to keep the lagrangian invariant. Hence this matrix version
of quantum group gauge theory also fails.  We note in passing that our
analysis does however, raise the interesting possibility of restoring a
kind of gauge symmetry {\em if we transform the $q$ also}, raising it
to a power with each gauge transformation. Another way out, which may
 be related to the recent preprint \cite{hirayama1}, is on each
transformation to use for $G$ the generators of another independent
commuting copy of $GL_q(2)$. Such an approach does not, however,
resemble any usual kind of gauge theory.

This problem with the Yang-Mills action based on a quantum trace
applies to both \cite{wu1} and \cite{isaev1}. Even if we can solve this
(perhaps by transforming $q$) we are still left with the problem of
justifying the desired transformation (\ref{curv.trans}) in terms for
the transformation of an underlying gauge potential $A_\mu$. This does
not seem likely. For example, it is attempted in \cite[Section
III]{wu1} and the following problem arises: each matrix component
$A_{ij}=dx^\mu A_\mu{}_{ij}$ is a one-form on $X$, yet the
transformation law proposed in \cite[eqn. (20)]{wu1} is
\eqn{conn.trans}{A'_{ij}=A_{kl}\tens (Su_{ik}) u_{lj}+1\tens
(Su_{ik})du_{kj}} where the quantum group matrix generators are denoted
$u_{ij}$ rather than $G_{ij}$ or $t^i{}_j$. The first part  is the quantum
adjoint action just as in (\ref{curv.trans}) written in a different way.
Unfortunately, the $d$ in the second part {\em is not} the space-time
$d$ but the bicovariant differential $d$ on the quantum gauge group. It
seems hard to interpret this then as some kind of
gauge theory
unless we identify the base and the quantum gauge group, which is not
possible in the setting of \cite{wu1} where the base is assumed
classical (its ring of functions $C(X)$ is commutative).

We see here the need for a fundamental ingredient in our quantum group
gauge theory of Section~2, namely in our gauge field transformation law
(\ref{beta.trans}),(\ref{betamat.trans}) the gauge transformation
maps $\gamma,\gamma^{-1}:A\to B$ take us from the quantum group $A$ to
$B=C(X)$ so that the $d$ in our $\gamma*d(\gamma^{-1})$ or $\gamma^i{}_k
d(\gamma^{-1})^k{}_j$ is truly the space-time differential as it should
be. This is also the role played in usual gauge theory by the gauge
transformation. It should be stressed also that in our formalism we do
not require $\gamma:A\to B$ to be an algebra map as this would surely be
too restrictive when $A$ is quantum and $B=C(X)$ is classical.

4. To conclude our discussion, we can summarise the situation as
follows:  we have obtained a working quantum group gauge theory on
classical spaces as a limit of the general theory in
\cite{brzezinski6}, which differs from \cite{wu1}\cite{isaev1} and
avoids problems present there. On the other hand, the formulae for the
gauge fields themselves in our approach (as obtained in Section 2) do
not make explicit use of the coalgebra structure $\Delta$. Yet it is
only here that the difference between $U_q(g)$ and $U(g)$ truly exists.
Thus the gauge group is a standard one. Moreover, even for non-standard
quantum groups $H$, the effective gauge group is a classical one in
this limit of \cite{brzezinski6}.

Let us note that this quantum group gauge theory on classical spacetime
can nevertheless differ in more subtle ways from the
usual one, when one considers matter fields.  If we describe these in
our original form in Section~1 as maps $\sigma:V\to C(X)$ while working
with $A_\mu(x),\gamma(x)\in H$ as in Section~2, the corresponding
formulae are as follows. Here $H$ is a quantum enveloping algebra such
as $U_q(g)$ and $V$ can be any algebra covariant under the
quantum group $H$ in the sense that it is a right $H$-module
algebra\cite[Sec. 6]{majid1}. It means that the product of $V$ is
covariant under an action $\ra$ (from the right) of $H$ in the sense
\eqn{modalg}{ (vw)\ra h=\sum (v\ra h\o)(w\ra h\t), \quad 1\ra
h=1\eps(h)\qquad\forall h\in H.} Thus, which algebras are covariant in
this way does depend on $\Delta$ and so does detect the difference between
$U_q(g)$ and $U(g)$. Since
the coproduct of $U_q(g)$ is not cocommutative, it tends to require
that $V$ is not commutative, i.e. not really the algebra of functions on any
actual
vector space.

For example, if $V$ is the Zamolodchikov algebra with generators $v^i$
as in Section 1, then a section $\sigma$ is equivalent to specifying
the $\C$-number fields \eqn{sigmafield}{ \sigma^{i_1\cdots
i_n}(x)=\sigma(v^{i_1}\cdots v^{i_n})(x); \quad n=1,2,\cdots} Since
$H$ acts from the right on the generators of $V$, it acts from the left
on these fields. Indeed, the transformation law (\ref{convtrans}) in Section~1
now reduces
to the assertion that these fields $\sigma^{i_1\cdots i_n}(x)$
transform in tensor products of the fundamental matrix representation
$\rho$ of $U_q(g)$ dual to $A(R)$,
\eqn{fieldtrans}{(\sigma^\gamma)^{i_1\cdots
i_n}(x)=<\gamma(x),t^{i_1}{}_{j_1}\cdots
t^{i_n}{}_{j_n}>\sigma^{j_i\cdots j_n}(x)=\rho^{\tens
n}(\gamma(x))^{i_1\cdots i_n}_{j_1\cdots j_n}\sigma^{j_1\cdots
j_n}(x).} Likewise, the covariant derivative comes out as
$\nabla=d+\rho^{\tens n}(A)$ as usual. We consider all the fields
together (for all $n$) because this is equivalent to working with the
entire (infinite-dimensional) quantum algebra $V$. Of course, one can
specialise to individual submodules such as $\sigma^i$ in the
fundamental quantum group representation.

We can also describe the matter fields equivalently in the dualised
way as in Section 2. We continue with $A_\mu(x),\gamma(x)\in H$ as
there. Thus $\sigma(x)\in C$ where $C=V^*$ in terms of Section~1. Such
$C$ are required in our formalism to be left $H$-module
coalgebras\cite[Sec. 6]{majid1}. Explicitly, it means they are linear
spaces equipped with a coalgebra $\Delta_C:C\to C\tens C$, $\eps_C:C\to\C$
which is
covariant under an action $\la$ of $H$ in the sense
\eqn{modcoalg}{(h\o\la\tens h\t\la)\Delta_C c=\Delta_C(h\la c),\quad
\eps_C(h\la c)=\eps(h)\eps_C(c),\qquad\forall h\in H,\ c\in C.}
This is equivalent to the description with
$V$ but in a different notation. The covariant derivative and gauge
transformations in this language are simply \eqn{coaltrans}{
\sigma^\gamma(x)=\gamma(x)\la\sigma(x),\qquad
\nabla\sigma(x)=d\sigma(x)+A(x)\la\sigma(x).}

For example, we can take here $C=H$ (as a coalgebra with
$\Delta_C=\Delta$) and $\la$ the left action by multiplication. Thus,
$\sigma(x),\gamma(x),A_\mu(x)\in H$ and $\nabla=d+A$ where $A_\mu(x)$
acts by multiplication from the left in $H$. Likewise,
$\sigma^\gamma(x)=\gamma(x)\sigma(x)$.

For another example, we can take $C=\und H$ the braided group
associated to $H$\cite{majid3} when $H$ is a true quantum group (with
universal R-matrix), and we take for $\la$ the quantum adjoint action.
Here $\und H$ has the same algebra as $H$ but its comultiplication
$\Delta_C =\und\Delta $ is not the usual one, being instead modified in
just such a way as to be covariant under the quantum adjoint
action\cite{majid3}.  Thus, $\sigma(x)\in \und H$ and its
transformation law is \eqn{sigmaadj}{\sigma^\gamma(x)=\sum
\gamma(x)\o\sigma(x) S\gamma(x)\t.} Thus, in our quantum group gauge
theory a particle $\sigma$ in such a quantum adjoint representation
transforms differently from the gauge field and curvature (the latter
transform by point-wise conjugation as we have seen in Section~2).
These two types of adjoint action are the same in the classical case of
$H=U(g)$ but become distinct in the quantum case. It means that gauge
fields are not the same thing as adjoint-covector fields in our picture
for algebraic reasons, in addition to the familiar geometrical
differences between gauge fields and covectors fields. Note that there
is no trouble writing down covariant equations of motion for any matter
fields $\sigma$, such as \eqn{eqnmotion}{\nabla^\mu\nabla_\mu\sigma =0}
since $\nabla$ in (\ref{coaltrans}) is covariant. On the other hand,
finding a corresponding lagrangian, as well as lagrangians for
sigma-models, appears to be more problematic. In the case of the
quantum adjoint representation (\ref{sigmaadj}) we have of course at
our disposal the invariant quantum trace (\ref{trace}) as well a
covariant product and a covariant coproduct $\und\Delta$.

There are plenty of other possibilities for comodule algebras $C$ in
which the matter fields can take their values (such as the coadjoint
action on $C=A(R)$). In all these cases we see that the formulae as in
(\ref{coaltrans}) are fairly analogous to the usual formulae, with the
field $\sigma(x)$ transforming in a representation $\la$ of the quantum
group. The difference between $U_q(g)$ and $U(g)$ shows up in the
allowed coalgebra structure on $C$. In fact, this is not so important
in the first approximation, in the same way as the algebra structure on
$V$ in Section~1 does not enter directly into the transformation
formulae. It is, however, needed for a correct quantum-geometrical
interpretation in the general theory\cite{brzezinski6} (where $V$ is a
quantum vector space). Of the same status are normalization conditions
corresponding to those mentioned in Section~1, which become now
$\eps(\gamma(x))=1$, $\eps(A_\mu(x))=0$ and $\eps_C(\sigma(x))=1$.
Again, they are not too essential but are needed for the correct
interpretation in \cite{brzezinski6}. For example, the last of these
says that $\sigma$ in (\ref{sigmaadj}) should be thought of as like a
group-valued field, while $\sigma-1$ is more like a Lie-algebra valued
field. One can also use the normalization $\eps_C(\sigma(x))=\lambda$ where
$\lambda$ is a constant (which could be zero) and thereby cover both cases.

A further place in this picture where the deformation of the
comultiplication plays a significant role is in the non-commutativity
of tensor product of representations. Thus, while the representations
of $U_q(g)$ coincide with those of $U(g)$, their tensor products
coincide only up to a non-trivial isomorphism. Moreover, if $V$ and $W$
are two spaces of the representation of $U_q(g)$, $V\tens W$ and
$W\tens V$ are not related in the usual way but by a non-trivial
braiding $\Psi$.  Likewise for the braided tensor product\cite{majid3}
of quantum vector spaces in Section~1 or of the module coalgebras
above.  Thus one may expect then that the effects of $q$-deforming the
gauge group (with the space-time classical) will show up quite
generally when one considers the interaction between different fields,
in the form of braid statistics. These are topics for further work.

\end{document}